\documentclass{bmcart}

\usepackage{graphicx}
\usepackage{tabulary,xcolor}
\usepackage{amsfonts,amsmath,amssymb}
\usepackage[utf8]{inputenc} 

\def\includegraphics{}

\startlocaldefs
\endlocaldefs

\begin{document}

\begin{frontmatter}

\begin{fmbox}
\dochead{Research}


\title{Predicting Antimicrobial Resistance in the Intensive Care Unit}


\author[
   addressref={aff1,aff2},                   
  corref={aff1},                       
   email={wty@bu.edu}   
]{\inits{TW}\fnm{Taiyao} \snm{Wang}}
\author[
   addressref={aff1},
   email={kyle.hansen@philips.com}
]{\inits{KRH}\fnm{Kyle R.} \snm{Hansen}}
\author[
   addressref={aff1},
   email={Joshua.k.loving@gmail.com}
]{\inits{JL}\fnm{Joshua} \snm{Loving}}
\author[
   addressref={aff2,aff3},
   email={yannisp@bu.edu, sites.bu.edu/paschalidis}
]{\inits{ICP}\fnm{Ioannis
  Ch.} \snm{Paschalidis}}
\author[
   addressref={aff1},
   email={Helencvanaggelen@gmail.com}
]{\inits{HvA}\fnm{Helen van} \snm{Aggelen}}
\author[
   addressref={aff1},
   email={eran.simhon@philips.com}
]{\inits{ES}\fnm{Eran} \snm{Simhon}}

\address[id=aff1]{
  \orgname{Philips Research North America}, 
  \city{Cambridge, MA},                              
  \cny{USA}                                    
}
\address[id=aff2]{%
  \orgname{Center for Information and Systems Engineering, 
    Boston University},
  \city{Boston, MA},
  \cny{USA}
}
\address[id=aff3]{%
  \orgname{Department of Biomedical Engineering, Boston University},
  \city{Boston, MA},
  \cny{USA}
}


\end{fmbox}


\begin{abstractbox}

\begin{abstract} 
Antimicrobial resistance (AMR) is a risk for patients and a burden for the healthcare
system. However, AMR assays typically take several days. This study develops
predictive models for AMR based on easily available clinical and microbiological
predictors, including patient demographics, hospital stay data, diagnoses, clinical
features, and microbiological/antimicrobial characteristics and compares those models to a naive antibiogram based model using only microbiological/antimicrobial characteristics. The ability to predict
the resistance accurately prior to culturing could inform clinical decision-making
and shorten time to action. The machine learning algorithms employed here show
improved classification performance (area under the receiver operating
characteristic curve 0.88-0.89) versus the naive model (area under the receiver operating
characteristic curve 0.86) for 6 organisms and 10 antibiotics using the Philips
eICU Research Institute (eRI) database. This method can help guide antimicrobial
treatment, with the objective of improving patient outcomes and reducing the usage of
unnecessary or ineffective antibiotics.

\end{abstract}




\end{abstractbox}
%

\end{frontmatter}



\section{Introduction}
Healthcare-associated infections (HAI) affect patients in a hospital or other
healthcare facility, and are not present or incubating at the time of admission. They
also include infections acquired by patients in the hospital or facility that appear
after discharge, and occupational infections among staff.  The estimated incidence
rate in the U.S. was 4.5\% in 2002, corresponding to 9.3 infections per 1000
patient-days and 1.7 million affected patients \cite{world2016health}.
It was estimated that there were 648,000 patients with 721,800 HAIs in U.S. acute care hospitals in 2011 \cite{magill2014multistate}.
According to the Centers for Disease Control and Prevention (CDC), there were an estimated 687,000 HAIs in U.S. acute care hospitals in 2015. About 72,000 hospital patients with HAIs died during their hospitalizations \cite{cdc2018}.
%
%

Antimicrobial resistance (AMR) is a growing threat to global health. In 2013, 
the CDC
reported that, each year, at least 2
million people become infected with antibiotic resistant bacteria in the U.S., and at
least 23,000 people die each year as a direct result of these infections \cite{cdc2013}. 
Although not all HAIs are caused by
resistant bacteria, antibiotic resistance is a great concern for hospitals, since
resistant pathogens can spread between patients and healthcare staff when hygiene
measures are insufficient and suboptimal antibiotic treatment can promote resistance,
thereby worsening patient outcomes. One of the main challenges facing clinicians is
the absence of fast and accurate antimicrobial resistance typing. Antimicrobial
resistance assays typically take several days to complete, but patients typically
require antimicrobial treatment the day they are admitted.

%
%
Being able to predict antimicrobial resistance accurately prior to culture-based resistance typing could thus help clinicians make an informed treatment decision in a timely manner. 

Machine learning has been proposed as a feasible solution for bacterial AMR
prediction. The current body of work on machine learning models for AMR prediction is
largely focused on {\em genomic data models}.  \cite{macesic2017machine} provides a
brief overview of current studies using machine learning for prediction of
antimicrobial susceptibility phenotypes from genotypic data. To the best of our
knowledge, all published predictions of AMR using machine learning algorithms use
either a k-mer representation of the bacterial genome
\cite{drouin2016predictive,santerre2016machine,davis2016antimicrobial,jha2017interpretable}
or other gene-related information
\cite{pesesky2016evaluation,rishishwar2013genome,her2018pan}. The model by Rishishwar
et al.~\cite{rishishwar2013genome} that discriminates between vancomycin-intermediate
and vancomycin-susceptible Staphylococcus aureus using 25 whole-genome sequences
reached an accuracy of 84\%. In a study by Pesesky et
al.~\cite{pesesky2016evaluation}, a rules-based and logistic regression prediction
achieved agreement with standard-of-care phenotypic diagnostics of 89.0\% and 90.3\%,
respectively, for whole-genome sequence data from 78 clinical Enterobacteriaceae
isolates. In a study by Drouin et al.~\cite{drouin2016predictive}, average test set
error rates of set-covering machine models that predict the antibiotic resistance of
Clostridium difficile, Mycobacterium tuberculosis, Pseudomonas aeruginosa, and
Streptococcus pneumoniae with hundreds of genomes and k-mer representation ranged
from 1.1\% to 31.8\%.  Davis et al.~\cite{davis2016antimicrobial} built adaptive
boosting classifiers with at least 100 genomes and k-mer representations to identify
carbapenem resistance in Acinetobacter baumannii, methicillin resistance in
Staphylococcus aureus, and beta-lactam and co-trimoxazole resistance in Streptococcus
pneumoniae with accuracies ranging from 88\% to 99\%. Her et al.~\cite{her2018pan}
used Support Vector Machine (SVM) algorithms with radial basis function kernels to
predict Escherichia coli AMR activities and reported Area Under the ROC Curve (AUC)
from 93\% to 100\% for 12 of the most-annotated antibiotics based on a pan-genome and
gene clusters selected by a genetic algorithm.

The primary objective in this work is to develop AMR predictive models and to
identify important variables to predict AMR {\em in the absence of genomic
  information}. Instead, we will leverage patient information and microbiology test
characteristics. Such models can be developed based on data that are more readily
available and easier to acquire and, as a result, have the potential to offer an
attractive alternative to models based on gene-related information.

Current practice relies on clinician interpretation of hospital antibiograms to guide prescribing of drugs based on population resistance. Hospital antibiograms summarize the percent of individual pathogens resistant to different antimicrobial agents and are derived from resistance typing results alone. A previous study estimated that standard care procedures result in appropriate prescription of antibiotics for only about 70\% of cases \cite{fleming2016prevalence}. Here we compare the performance of naive models based only on microbiology data (analogous to an antibiogram) with predictive models that combine resistance typing with information about patient demographics, hospital stay, previous resistance test results, and diagnoses.  

An important enabler of our work is the increasing availability of patients’ Electronic Health Records.  The digitization of patients’ medical records over the
last two decades has enabled the development of more personalized and accurate models
for diagnostics and treatment. Over the past few years, there has been increased
interest in algorithmic and data-driven approaches to improve healthcare quality.
Machine learning is increasingly being used with Electronic Health Records to predict
chronic disease hospitalizations \cite{brisimi2018predicting,ACC-diab-bri-xu-2019},
level-of-care requirements\cite{hao2020early}, 
mortality\cite{wang2020predictive,chiang2019risk} and readmission \cite{badawi2012readmissions,wang2019prescriptive,bertsimas2020prescriptive,wang2020data}.
%
%
%

The remainder of this paper is organized as follows.  In Section \ref{sec:methods},
we review the database specifics, data selection, feature generation, pre-processing
and classification methods used in the study.  In Section~\ref{sec:results}, we
present the prediction results and highlight the important variables. In
Section~\ref{sec:discussion}, we discuss the results, including both limitations and
strengths. Conclusions are in Section~\ref{sec:conclusions}.

\section{Methods} \label{sec:methods}
\subsection*{Settings and patient data}
In this study we utilized microbiology tests and patient information from the eICU
Research Institute (eRI) database for patients who had a complete hospitalization
between January 1, 2007 and March 31, 2013. Detailed descriptions of the eRI database
are provided in \cite{lilly2011benchmark,mcshea2010eicu}. We filtered tests such that
patients were at least $16$ years old and cultures were taken from patients while
being admitted to an ICU unit.

We selected microbiology tests limited to 6 organisms and 10 antibiotics.  Organisms
included are Staphylococcus aureus, Escherichia coli, Klebsiella pneumoniae,
Pseudomonas aeruginosa, Staphylococcus epidermidis, Enterobacter cloacae and
antibiotics were one of vancomycin, imipenem/cilastatin, cefipime, oxacillin,
ciprofloxacin, nitrofurantoin, trimethoprim/sulfamethoxazole, cefazolin,
ampicillin/sulbactam, and ampicillin.

After selection, there were 12,575 unique patients, 13,087 unique patient unit stays,
and 80,125 tests from 191 different units. The test results which we aimed to predict
were distributed as follows: 55,095 were Sensitive, 22,852 were Resistant, and 2,178
were Intermediate. Thus, after 'Sensitive' and 'Intermediate' records were combined
as 'non-Resistant', the average AMR rate was 28.5\%.

We included in the study two types of variables as described below.
\begin{itemize}
\item Patient data: patient unit stay id, gender, age, ethnicity, height, admission
  weight, unit location id, unit type, unit stay type, unit admit source, unit admit
  time, hospital admit source, number of minutes from unit admit time to culture
  time, the ICU visit number during the patient’s hospital stay and admission
  diagnosis for patient unit stay.

\item Microbiology data: patient culture taken year, culture taken time, number of
  minutes from unit admit time until the culture was taken, culture site, organism,
  antibiotic, and sensitivity level of antibiotic.
\end{itemize}

\subsection*{Feature generation and pre-processing}
The feature generation and pre-processing consisted of the following steps: 
\begin{itemize}
\item Generate feature interactions `anti-organism' between `organism' and
  `antibiotic'. 
\item Generate existing resistant test information for the same patient and same type
  `anti-organism' (more than 48 hours ago). 
\item Transform hospital admit time and culture taken time from minutes to days and
  their log transformation. 
\item Convert categorical variable into dummy/indicator variables with one-hot
  encoding (e.g., `anti-organism', `locationid', `apacheadmissiondx').  
\item Cutoff lower-tail 
($0.5$th percentile) and upper-tail ($99.5$th percentile) values for height, weight, hospital admit time
  and culture taken time.  
\item Replace missing values of admit weight by discharge weight and replace other
  variables' missing values by the median across all patients. 

\item Perform feature selection by a two-sided $t$-test which compares the means of a
  variable in the 'resistant' and 'non-resistant' cohorts. We take as 'null'
  hypothesis that means are equal. If the corresponding $p$-value is below a
  threshold of 0.1
  , the variable is retained. We also perform an additional step of
  feature elimination, excluding one variable among each set of two highly correlated ($>0.75$ or $<-0.75$)
  variables.
\item Transform features by scaling each feature to be between zero and one (i.e., by
  dividing by the total range of the feature). 
\end{itemize}


Microbiology tests were randomly split by patient unit stay id in a 60\%:20\%:20\%
fashion to training, test and validation cohorts.  

\subsection*{Machine learning methods}
We tested a variety of supervised classification methods \cite{276176:6227309},
including $\ell_1$-regularized logistic regression (L1LR), random forests (RF),
neural networks (NN) and Gradient Boosting Machine (GBM). All methods were
implemented in Python, using scikit-learn\unskip~\cite{276176:6227142} and
LightGBM\unskip~\cite{276176:6227141}.  We compared these methods with a naive model (AB) based on an antibiogram by calculating the percentage resistance in the training and validation datasets and naively setting that equal to the probability of resistance in the test datasets. 

$\ell_1$-regularization can mitigate overfitting and improve the interpretability for
models in clinical settings
\cite{brisimi2018predicting,ACC-diab-bri-xu-2019,chen2018robust,xu2016joint,wang2019convergence}.
Logistic regression, widely used in statistics and machine learning studies, was
implemented with an $\ell_1$-regularization term to induce sparsity and
interpretability. We tuned the strength of the regularizer using cross-validation. 

The random forest (RF) method builds a large collection of de-correlated trees, and then
averages them, which generally leads to a substantial performance improvement over
single tree classifiers. We tuned a number of hyperparameters, such as the number of
trees, the depth of each tree, and the number of features used (a random subset of the
total) at each tree node split. 

Gradient boosting machine (GBM), also referred to as gradient-boosted decision trees,
is a popular machine-learning algorithm used for regression and classification
tasks. We used LightGBM which is a fast and high-performance GBM framework that grows
trees leaf-wise rather than level-wise and implements a host of techniques, such as
gradient-based one-side sampling and exclusive feature bundling to deal with a large
number of data instances and features, respectively. We tuned hyperparameters such as
the number of leaves in each tree, and the minimum number of sample points
corresponding to each leaf.

Finally, we used a class of feed-forward artificial neural networks (NN), called a
multilayer perception, which consists of at least three layers of nodes. Except for
the input nodes, each node uses a nonlinear activation function. It can distinguish
data that is not linearly separable due to its multiple layers and non-linear
activation. We tuned a number of hyperparameters such as the number of hidden
neurons, layers, and iterations using a rectified linear unit (`relu') activation
function for each hidden layer, and the 'Adam' adaptive stepsize rule for stochastic
gradient descent.

A Receiver Operating Characteristic (ROC) curve is created by plotting the true
positive rate (or recall, or sensitivity) against the false positive rate (equal to
one minus the specificity) at various thresholds. The $c$-statistic or the area under
the ROC curve (AUC), is used to evaluate the prediction performance.  A perfect
predictor gives an AUC score of 1 and a predictor which makes random guesses has an
AUC score of 0.5.  

In machine learning, ensemble methods\unskip~\cite{276176:6227309} that use multiple
learning algorithms can obtain better predictive performance than any learning
algorithms alone. We proposed a new ensemble model by averaging other prediction
results based on neural networks, GBM and RF, which achieved the highest AUC in
the eRI dataset.

\section{Results} \label{sec:results}
\subsection*{Prediction accuracy and ROC curve}
We trained each model by using 80\% of the samples for training model parameters
(60\% training, 20\% validation) and evaluated the model’s performance on the unseen
test 20\%. 

Table \ref{table-AUC-all} summarizes the performance of the various predictive models
on the test dataset under two learning modes. The first column contains the names of
machine learning methods.  The second column is the AUC under the assumption that no
temporal effects exist, in which case training and test datasets are formed by
randomly selecting patients. The third column is the AUC under the assumption that
temporal effects exist; in this mode, samples are split by unit admit year and unit
stay id so that the training samples belong to patients hospitalized before a certain
data and test samples correspond to patients after that date. The latter mode aims at
capturing how a predictive model would be trained in practice. 
From the comparison of the AUC column and the AUC\_time column, the prediction
accuracy in practice may decrease due to temporal trends. In particular, changes in
data collection and prevalence of specific microbes, render earlier data less
predictive about the future.

\begin{table}[!htbp]
\caption{{ Prediction accuracy by different methods.} }
\label{table-AUC-all}
\def\arraystretch{1}
\ignorespaces 
\centering 
\begin{tabulary}{\linewidth}{LLL}
\hline 
  Methods  & AUC     & AUC\_time \\
    \hline 
AB       & 86.01\% & 83.03\%   \\
L1LR     & 88.93\% & 86.47\%   \\
RF       & 89.03\% & 86.55\%   \\
NN       & 88.44\% & 86.34\%   \\
GBM      & 89.26\% & 86.91\%   \\
ensemble & 89.46\% & 87.13\%   \\
\hline
\end{tabulary}\par 
\end{table}

Figure 1 
are the ROC curves for the AMR prediction models and the naive model (AB) in the test cohort.
Figure 2 
shows the ROC curves for the AMR prediction models and the naive model (AB) in the test cohort  under the assumption that
temporal effects exist.


In order to study the difference of prediction accuracy on different organisms, we
trained the L1LR model for every organism separately by using an 80\%-20\% split
between training and testing data subsets. Table~\ref{table-AUC6-L1LR} summarizes the
performance of the L1LR predictive models 
versus the naive model (AB)
on the test dataset for every organism.
\begin{table}[!htbp]
\caption{{ Prediction accuracy for 6 organisms by L1LR.} }
\label{table-AUC6-L1LR}
\def\arraystretch{1}
\ignorespaces 
\centering 
\begin{tabular}{p{0.32\linewidth}p{0.14\linewidth}p{0.14\linewidth}p{0.14\linewidth}}
\hline 
 &
  AUC &
  AUC AB &
  sample size
   \\
  \hline 
Enterobacter cloacae &
  93.66\% &
  92.03\% &
  4198 \\
Staphylococcus epidermidis &
  92.90\% &
  91.35\% &
  4289 \\
Staphylococcus aureus &
  91.48\% &
  84.22\% & 
  27733 \\
Klebsiella pneumoniae &
  89.96\% &
  80.21\% &
  12814 \\
Pseudomonas aeruginosa &
  85.81\% &
  80.30\% &
  9382 \\
Escherichia coli &
  80.52\% &
  76.97\% &
  21709 \\
\hline
\end{tabular}\par 
\end{table}

\subsection*{Important variables}
Table \ref{table-imp} summarizes the 20 most predictive variables and coefficients in
the L1LR model.  Most of them are binary variables with prefix `ao' from one-hot
encoding of feature interactions `anti-organism' between `organism' and `antibiotic'.
Others include y\_pre, i.e., the mean of existing resistant tests for the same
patient and same type `anti-organism' more than 48 hours ago, admission diagnosis
with prefix `apacheadmissiondx', unit location id with prefix `locationid', and
culture taken time from ICU admit time. We note that since the variables have been
standardized, the coefficients of different variables reveal relative strength of the
variable in the model.

Figure 3
 summarizes average AMR rates with 6 organisms and 10 antibiotics.
Figure 4
 summarizes total frequency counts. 
Figure 5
 summarizes AMR frequency counts.

\begin{table}[!htbp]
\caption{{Variable coefficients in the L1LR model.} }
\label{table-imp}
\def\arraystretch{1}
\ignorespaces 
\centering 
\footnotesize
\begin{tabular}{p{0.85\linewidth}p{0.1\linewidth}}
\hline 
L1LR                                                                  & coef. \\
\hline 
ao\_nitrofurantoin\_Staphylococcus aureus                 & -5.88 \\
ao\_vancomycin\_Staphylococcus epidermidis                & -5.73 \\
ao\_imipenem/cilastatin\_Escherichia coli                 & -5.51 \\
ao\_vancomycin\_Staphylococcus aureus                     & -4.90 \\
ao\_cefazolin\_Enterobacter cloacae                       & 4.69  \\
ao\_ampicillin\_Pseudomonas aeruginosa                    & 4.59  \\
y\_pre                                                    & 4.54  \\
ao\_ampicillin\_Klebsiella pneumoniae                     & 4.51  \\
ao\_ampicillin/sulbactam\_Pseudomonas aeruginosa          & 4.18  \\
ao\_nitrofurantoin\_Pseudomonas aeruginosa                & 3.99  \\
ao\_imipenem/cilastatin\_Klebsiella pneumoniae            & -3.30 \\
apacheadmissiondx\_Ventriculostomy                        & -3.27 \\
ao\_nitrofurantoin\_Staphylococcus epidermidis            & -3.23 \\
ao\_trimethoprim/sulfamethoxazole\_Pseudomonas aeruginosa & 3.06  \\
ao\_cefazolin\_Pseudomonas aeruginosa                     & 3.03  \\
ao\_imipenem/cilastatin\_Enterobacter cloacae             & -2.97 \\
apacheadmissiondx\_Spinal cord surgery, other             & -2.96 \\
ao\_nitrofurantoin\_Escherichia coli                      & -2.90 \\
ao\_ampicillin\_Enterobacter cloacae                      & 2.89  \\
ao\_ampicillin\_Staphylococcus epidermidis                & 2.76 \\
     \hline 
\end{tabular}
\end{table}

%
%



\section{Discussion} \label{sec:discussion}
In this study, we apply a set of machine learning methods and propose a new ensemble
method to predict AMR utilizing electronic medical records (EMR) data in conjunction with microbial information. We show that it is an improvement over a naive model (AB) designed to represent an antibiogram, which is the common practice for antibiotic resistance prediction in hospitals. The results show strong classification performance using the
eRI database.  In addition, important variables have been identified offering
interpretability of the results.

This study provides a new framework and new predictions using EMR for predicting AMR.  
Instead of focusing on one organism as in previous studies, we
explore six organisms and ten antibiotics.  To our knowledge, this is the first study
with high accuracy in predicting AMR and identifying important variables without
gene-related information.

%
%

The results have important consequences for ICU patients.  In an ICU setting, it is
critically important to obtain quick and accurate prediction without DNA sequencing,
since about half of the ICU patients with AMR tests in the study are from the
emergency room.  The time-frame for the detection is also drastically shorter than
testing multiple antibiotics -- a common practice. Early detection and treatment of
HAIs are vital for reducing the length of stay and
mortality.
For ICU patients
who need acute treatment, 
%
%
other researchers and hospitals can use our framework with other datasets.

Researchers can further combine the variables used here with the genomic information that may improve the prediction performance.

The most important variables are interaction features `anti-organism' between `organism' and `antibiotic'.
If only `organism' and `antibiotic' are used without the second-order interaction
terms (e.g., ao vancomycin Staphylococcus epidermidis), tree-based models can learn
the interaction relationship but the interpretable logistic regression cannot so it
performs more poorly without computing those interaction terms.
%
%
%
%

Other important variables are existing AMR tests for the same patient and same type
`anti-organism', which are highly correlated if the culture taken times are close.
However, in clinical settings, it takes 24--48 hours to obtain resistant test
results.  That is why only existing resistance tests more than 48 hours ago should be
considered as variables in the models.  If existing resistance tests more than 48
hours ago are removed, values in the AUC
column of Table \ref{table-AUC-all}
decrease by about 1\%.  Average AMR rates and AMR frequency counts with 6 organisms
and 10 antibiotics in Figure 3
 and Figure 5
 can be used to help select antibiotics and reduce drug
misuse for patients.  For instance, if the average AMR rate is higher than 95\% for
one `anti-organism', e.g., Klebsiella pneumoniae and ampicillin, the antibiotic drug
should not be recommended for the treatment of bacterial infections.  However, if the
number of total frequency counts in Figure 4
 is small, the AMR
results should not be trusted.
Other features like weight, age, height and ethnicity are important in tree-based models, e.g., GBM.

In order to study AMR temporal effects, we have carried out an experiment in which
samples are split by unit admit year and unit stay id so that the training samples
belong to patients hospitalized earlier and the test samples belong to the patients
hospitalized after a certain cutoff date. Such a prediction seeks to reproduce how a
predictive model would be used in practice.  From the comparison of the AUC column
and the AUC\_time column in 
Table \ref{table-AUC-all}, the prediction accuracy in
practice may decrease due to temporal trends, potentially because the type of AMR
instances evolves over time. 
This can also be observed in the comparable shift in AUC values between Figure 1 
 and Figure 2.
As Figure 6 
 suggests, average AMR
rates change over time for some organisms and antibiotics.

There are some limitations of our study.  First, spatio-temporal effects
\cite{mccormick2003geographic} are not explicitly modeled. One reason is that our
data span more than 6 years and 191 different units; many more data samples would be
needed to reliably model spatio-temporal effects.  
    
\section{Conclusions} \label{sec:conclusions}
The machine learning algorithms employed in this work to predict AMR based on the eRI
database exhibit improved classification performance compared to the AB model.  Our framework is faster and less
expensive than previous research on predicting AMR with genomic data. The techniques we utilized can be extended to other settings besides the
ICU. Our findings and the development of accurate predictive models of AMR can guide
the choice of appropriate antibiotics, thereby reducing the occurrence of antibiotic
resistance, decreasing healthcare costs and improving patient outcomes.

The models developed in this work are based on patients who are admitted to the ICU at
least once. In the future, the models can be further generalized to other hospital
units, evaluating their performance in a more general setting. Another future direction is to
combine our methods with previous predictions that leveraged gene information. As we
argued, our framework is faster and more inexpensive but perhaps less accurate than
genomics-based models. For some organisms or `anti-organism' combinations, our
predictions achieve high prediction accuracy. But for some other organisms or
`anti-organism' combinations with low prediction accuracy, e.g., `Escherichia coli',
one could potentially refine the prediction with DNA sequencing
data~\cite{her2018pan}. The models introduced here can help hospitals speed up
appropriate treatment of infection to improve patient outcomes, and to reduce the
costs of AMR prediction.

\begin{backmatter}

\section*{Abbreviations}
AMR:
Antimicrobial resistance

eRI:
eICU Research Institute

APACHE:
Acute Physiology and Chronic Health Evaluation

AUC:
Area Under the Receiver Operating Characteristic (ROC) Curve

CDC:
Centers for Disease Control and Prevention 

EMR:
electronic medical records 

GBM:
Gradient Boosting Machine 

HAI:
Healthcare-associated infection

ICU:
Intensive care unit

L1LR:
$\ell_1$-regularized logistic regression

NN:
neural networks 

RF:
random forests 
 
SVM:
Support Vector Machine 

\section*{Competing interests}
ES, JL, KRH and HvA were salaried employees of Philips Research when the study was done. ICP and TW declare no conflicts of interest. 
ICP has been partially supported by the Office of Naval Research under MURI
        grant N00014-19-1-2571, by the NSF under grants
        IIS-1914792, DMS-1664644, and CNS-1645681, and by the NIH/NIGMS under grant
        1R01GM135930.
\section*{Author's contributions}
All of us contributed to the design of the study and to the definition of the problem. TW and KRH queried and analyzed the data.
TW took the lead in writing the manuscript. All authors
provided critical feedback and helped shape the research, analysis and manuscript.

\section*{Acknowledgements}
The authors would like to thank Chieh Lo, Ting Feng, Bryan Conroy, David Noren, and
Andrew Hoss for helpful discussions. 


\bibliographystyle{bmc-mathphys} 
\bibliography{bmc_article}      




\section*{Figures}
  \begin{figure}[h!]
  \caption{\csentence{ROC curves for the AMR prediction \& naive models in the test cohort.}
      }
      \end{figure}

    \begin{figure}[h!]
        \caption{\csentence{ROC curves for the AMR prediction \& naive models in the test cohort under the assumption that
    temporal effects exist.}
        }
    \end{figure}

    \begin{figure}[h!]
        \caption{\csentence{Percentage of AMR averaged across all patients and samples for the selected 6 organisms and 10 antibiotics.}
        }
    \end{figure}
    \begin{figure}[h!]
        \caption{\csentence{Total frequency counts with 6 organisms and 10 antibiotics.}
        }
    \end{figure}
    \begin{figure}[h!]
        \caption{\csentence{AMR frequency counts with 6 organisms and 10 antibiotics.}
        }
    \end{figure}
    
    \begin{figure}[h!]
        \caption{\csentence{Average AMR rates over years for high-frequency organisms and antibiotics.}
        }
    \end{figure}





\end{backmatter}
\end{document}